\documentclass[twocolumn,english,prx,showpacs,longbibliography]{revtex4-1}

\usepackage{amssymb}
\usepackage{graphicx}
\usepackage{dcolumn}
\usepackage{bm}
\usepackage{amsmath}

\def\OMIT#1 {{}}
\def\MEMO#1 {{}}

\newcommand{\sjtu} {Department of Physics and Astronomy, Shanghai Jiao Tong University, Shanghai 200240, People's Republic of China}
\newcommand{\lmu} {Department of Physics and Arnold Sommerfeld Center for Theoretical Physics,
Ludwig-Maximilians-Universit{\"a}t M{\"u}nchen, Theresienstr.\ 37,
80333 Munich, Germany}
\newcommand{\iqoqi} {Institute for Quantum Optics and Quantum Information, Austrian Academy of Sciences, 6020 Innsbruck, Austria}

\begin{document}
\title{Universal long-time behavior of stochastically driven interacting quantum systems}

\author{Zi Cai}
\affiliation{\sjtu}
\affiliation{\iqoqi}
\author{Claudius Hubig}
\affiliation{\lmu}
\author{Ulrich Schollw\"{o}ck}
\affiliation{\lmu}

\date{ \today }

\begin{abstract}
One of the most important concepts in non-equilibrium physics is relaxation. In the vicinity of a classical critical point, the relaxation time can diverge and result in a universal power-law for the  relaxation dynamics; the emerging classification of the dynamic universality class has been elucidated by Hohenberg and Halperin. In this paper, we systematically study the long-time relaxation dynamics in stochastically driven interacting quantum systems. We find that even though the stochastic forces will inevitably drive the systems into a featureless infinite temperature state, the way to approach the steady state can be highly nontrivial and exhibit rich universal dynamical behavior determined by the interplay between the stochastic driving and quantum many-body effects. We investigate the dynamical universality class by including different types of perturbations. The heating dynamics of a Hamiltonian with locally unbounded Hilbert space is also studied.
\end{abstract}

\maketitle

\section{Introduction}

Recently, non-equilibrium quantum many body physics has attracted considerable attention due to enormous progress in ultracold atomic experiments\cite{Eisert2015}, in which interacting quantum systems can be driven out of thermodynamic equilibrium by quenching\cite{Rigol2008,Polkovnikov2011,Gring2012,Trotzky2012,Alessio2015}, ramping\cite{Braun2015} and periodic driving\cite{Struck2011,Struck2012,Jotzu2014} of Hamiltonian parameters, or by coupling the systems to engineered\cite{Diehl2008,Verstraete2009,Daley2009,Barreiro2011,Diehl2011b} or thermal\cite{Cazalilla2006,Prosen2008,Prosen2011b,Horstmann2013,Rancon2013,Cai2014,Znidaric2015} baths and external  noise\cite{Torre2010,Marino2012,Poletti2012,Cai2013,Sieberer2013,Buchhold2015,Marino2016,Marino2016a}.  Unlike the Boltzmann distribution in equilibrium physics, the derivation of a universal distribution for the non-equilibrium systems in phase (Hilbert) space is an essential theoretical challenge. Incorporating quantum many-body effects further complicates the systems and gives rise to important novel phenomena, {\it e.g.} a system subject to external driving forces can display unexpected dynamical and steady properties absent in the non-driven counterpart\cite{Kitagawa2010,Lindner2011,Jiang2011,Hauke2012,Rudner2013,Goldman2014,Goldman2015}.  Up to now, most of the research on driven systems has been devoted to the periodically driven cases\cite{Russomanno2012,Alessio2014,Lazarides2014a,Chandran2015,Citro2015,Ponte2015,Weidinger2016}, while their stochastic counterpart is much less studied. However, understanding stochastically driven quantum (many-body) systems is not only of high theoretical interest, but also of immense practical significance due to the possible relations to the decoherence problem  in quantum simulation and information processing\cite{Hu2015}.

Even though the final state of periodical driven system is still a subject of debate \cite{Alessio2014,Lazarides2014a,Chandran2015,Citro2015}, stochastic driving of a quantum system will inevitably lead to decoherence and heat the system towards an infinite-temperature state, irrespective of the specific forms of the Hamiltonian and the driving. In spite of the triviality of the steady state, the way to approach the steady state can exhibit rich dynamical behavior. For example, the relaxation rate towards the steady state may diverge for certain types of many-body systems and dynamical properties can exhibit some forms of universal power-law behavior that are robust under some perturbations while sensitive to others. This abnormal relaxation behavior is strongly reminiscent of the \textquotedblleft critical slowing down\textquotedblright\ phenomena in classical systems: in the vicinity of a classical critical point, not only the static correlations, but also the dynamical properties exhibit some universal power-laws. The emerging classification of the dynamic universality class was first elucidated by Hohenberg and Halperin\cite{Hohenberg1977} and has become an important branch of modern statistical physics. Back to quantum many-body physics, it is natural to ask whether we can find similar universal power-law relaxation dynamics, and, if so, what universality class they belong to  and what kind of perturbations are relevant (irrelevant) for the asymptotic long-time behavior.

Rather than trying to find general answers, we address the above questions for specific examples of quantum many-body systems with one of the simplest stochastic driving protocols: a telegraph-like driving where one of the Hamiltonian parameters randomly jumps between two discrete values during the time evolution. This type of noise has recently been introduced to condensed matter physics for the first time to model the noise effect on the intensively studied Majorana fermions\cite{Hu2015}.  The goal in this paper is to understand the statistical long-time behavior of many-body quantum systems driven by a stochastic  sequence of  sudden quenches, and its dependence on the specific form of the Hamiltonian and various perturbations, especially the interactions.   To do that, we shall explore two different examples: a one-dimensional (1D) spinless fermonic model and a two-dimensional (2D) bosonic quantum $O(N)$ model, representing the Hamiltonians with locally bounded and unbounded Hibert space respectively. By exploring the  long-time behavior in these specific models, we have taken a first step towards understanding the stochastically driven interacting quantum systems.

The paper is organized as follows: in Sec.~\ref{sec:Methods}, we propose a general method to deal with the quantum many-body systems with telegraph-like stochastic driving, and discuss its limitations;  in Sec.~\ref{sec:1Dfermion}, we investigate  a 1D spinless fermonic model with a stochastically fluctuating staggered potential, which exhibits algebraic relaxation dynamics. We further examine the effect of three common perturbations (pairing, disorder, and especially an interaction) on this long-time behavior. In Sec.~\ref{sec:ON}, we study a 2D quantum $O(N)$ model with a fluctuating mass that drives the system crossing the phase boundary of the equilibrium phase diagram. At least in the large-$N$ limit, we find that the interaction will significantly suppress the heating dynamics and change the divergent behavior from exponential to algebraic. Sec.~\ref{sec:experiment} and  Sec.\ref{sec:conclusion} contain a discussion of possible experimental realizations and an outlook.

 \begin{figure}[htb]
\includegraphics[width=0.85\linewidth]{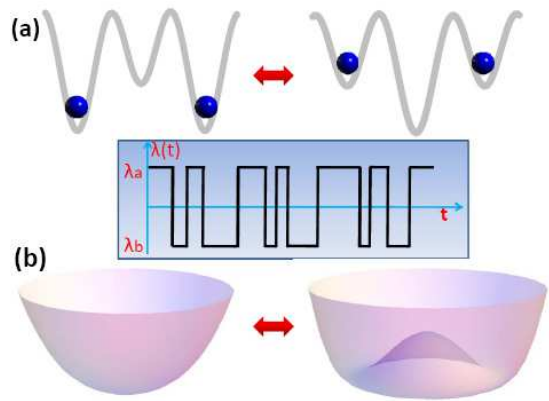}
\caption{Sketches of the two models studied in this paper: (a) a 1D spinless fermion model with fluctuating staggered potential and (b) a 2D quantum $O(N)$ model with a fluctuating mass. The inset is an example of a typical trajectory of the telegraph-like stochastic driving parameter $\lambda(t)$.}
\label{fig:fermion}
\end{figure}


\section{Marginal density matrix method}
\label{sec:Methods}

In this paper, we consider stochastically driven quantum many-body systems
in which one parameter $\lambda(t)$ in the Hamiltonian randomly jumps between two values $\lambda_a$ and $\lambda_b$ with a transition rate $\kappa$ during time evolution. For a given trajectory of the parameter $\{ \lambda(t)\}$ (e.g the inset of Fig.1), the system evolves unitarily with a time-dependent Hamiltonian and can be described by the density matrix $\rho_{\{\lambda(t)\}}(t)$. Since we are interested in the long-time behavior near the steady state (infinite-temperature state), we assume that the ergodic hypothesis holds, provided that the stochastic driving force has no long-range temporal correlation. As a consequence, for physical observables, the time average over a long period of time is equal to ensemble averages over all the stochastic trajectories.  Our goal is to derive an equation of motion (EOM) for the average density matrix  $\rho_s(t)=\langle \rho_{\{\lambda(t)\}}(t)\rangle_s$ where the angular brackets $\langle \quad \rangle_s$ denote the ensemble average over all stochastic trajectories. To achieve this goal, we introduce the marginal density matrix $\rho_{a(b)}(t)$ in which the ensemble average is over those trajectories satisfying $\lambda(t)=\lambda_{a(b)}$:
\begin{equation}
\rho_{a(b)}(t)=\langle \rho(t)\delta(\lambda(t)-\lambda_{a(b)})\rangle_s.
\end{equation}
Obviously $\rho_s(t)=\rho_a(t)+\rho_b(t)$. The average of physical observable is defined as $\langle \hat{O}\rangle= \text{Tr}(\hat{O}\rho_a)+\text{Tr}(\hat{O}\rho_b)$. The marginal density matrix method was first introduced by Zoller. {\it et al.} in the context of quantum optics\cite{Zoller1981}, and recently been introduced to condensed matter physics\cite{Hu2015}.  We can prove (see the Appendix and Ref.\cite{Hu2015})  that the EOM of the marginal density matrix is described by the following master equation:
 \begin{eqnarray}
 \nonumber \frac{d\rho_a(t)}{dt}=i[\rho_a,\hat{H}_a]-\kappa\rho_a+\kappa\rho_b\\
 \frac{d\rho_b(t)}{dt}=i[\rho_b,\hat{H}_b]+\kappa\rho_a-\kappa\rho_b  \label{eq:Master1}
 \end{eqnarray}
 where $\hat{H}_{a(b)}$ is the time-independent Hamiltonian for $\lambda(t)=\lambda_{a(b)}$. Assuming the dimension of the Hilbert space of the system is $\boldsymbol{N}$, we can rewrite the $\boldsymbol{N}\times \boldsymbol{N}$ density matrix $\rho_{a(b)}$ into a $ \boldsymbol{N}^2$-dimensional vector  $\vec{\rho}_{a(b)}$, and the master equation Eq.(\ref{eq:Master1}) turns to:
  \begin{equation}
  \frac{d\vec{\rho}_s}{dt}=\hat{\mathbb{L}}\vec{\rho}_s\label{eq:Master2}
   \end{equation}
  in which $\vec{\rho}_s=[\vec{\rho}_a,\vec{\rho}_b]^T$ is a $2\boldsymbol{N}^2$-dimensional vector and $\hat{\mathbb{L}}$ is the $2\boldsymbol{N}^2\times2\boldsymbol{N}^2$  Liouville superoperator defined in Eq.~(\ref{eq:Master1}). For a quadratic Hamiltonian with translational symmetry, we can perform a Fourier transformation $H_{a(b)}=\sum_k H^{a(b)}_k$, and the Liouville superoperator can be decomposed as $\mathbb{L}=\bigotimes_k \mathbb{L}_k$. In general, the long-time behavior of the system is determined by the spectrum (eigenvalues) of $\mathbb{L}_k$. From Eq.~(\ref{eq:Master1}), we can find that the steady state is always a unit matrix irrespective of the specific form of the Hamiltonian: $\rho^s_a=\rho_b^s=\hat{\mathbf{1}}/(2\boldsymbol{N})$, corresponding to the infinite-temperature state.

Before we proceed further to discuss specific examples, we make some remarks about the method. Compared to the conventional method of calculating unitary evolution for each given trajectories and then do the ensemble average,   the  marginal density matrix method has the advantage of the absence of stochasticity. The ensemble average has already been performed implicitly in Eq.~(\ref{eq:Master1}) with the price of the Hibert space being significantly enlarged from $\boldsymbol{N}$ to $2\boldsymbol{N}^2$. For a fermonic or bosonic quadratic Hamiltonian, the EOM of the system can be reduced to that of the single-particle correlation functions taking advantage of Wick's theorem; thus the dimension of the reduced EOM is proportional to the system size $L$. However, for a genuine interacting quantum many-body systems $\boldsymbol{N}\sim \mathcal{O}(e^{L})$, thus it is not convenient to directly solve Eq.~(\ref{eq:Master1}) for large systems.

\begin{figure*}[htb]
\includegraphics[width=0.325\linewidth,bb=23 23 286 210]{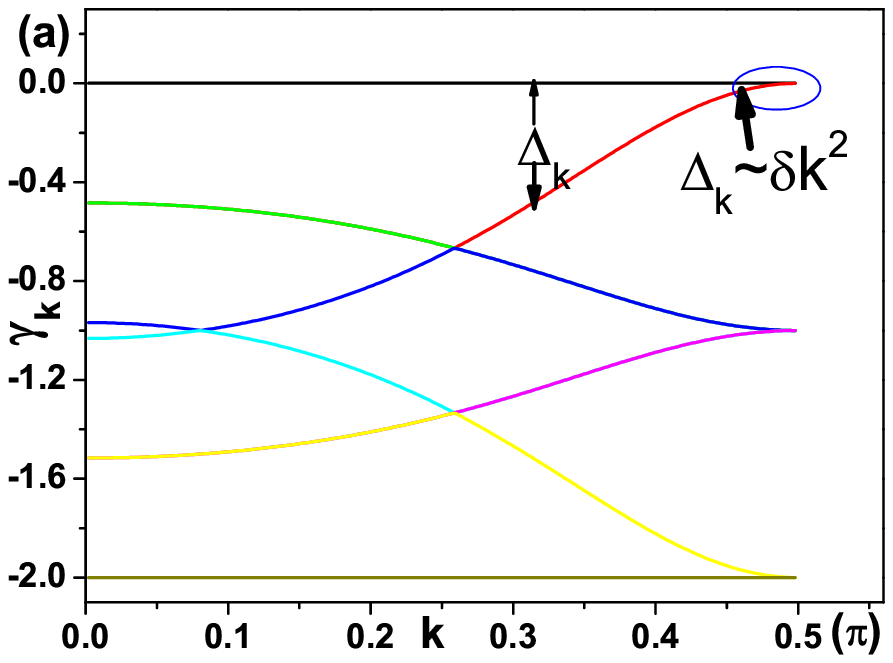}
\includegraphics[width=0.325\linewidth,bb=18 24 275 215]{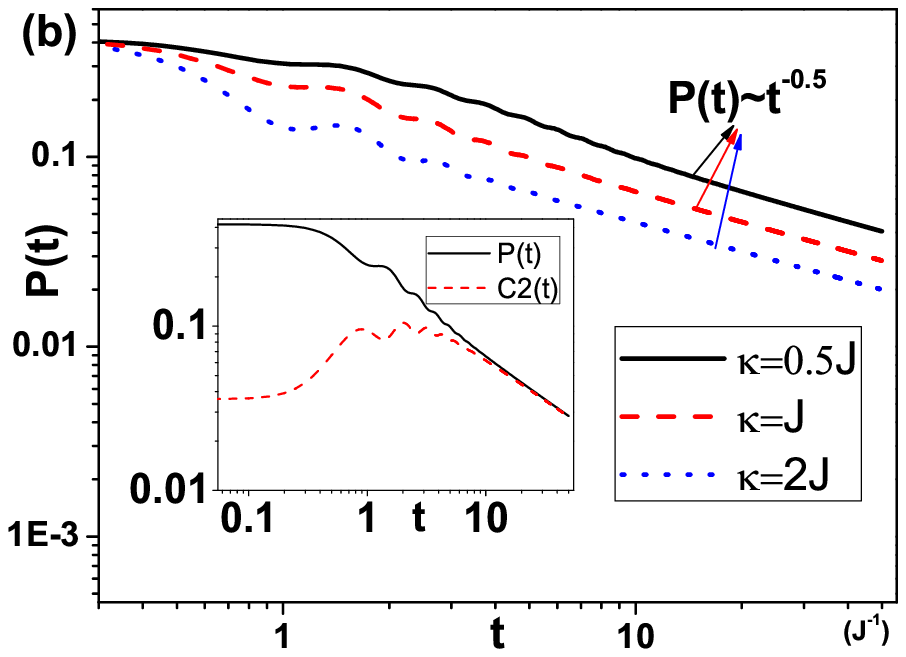}
\includegraphics[width=0.325\linewidth,bb=18 24 275 210]{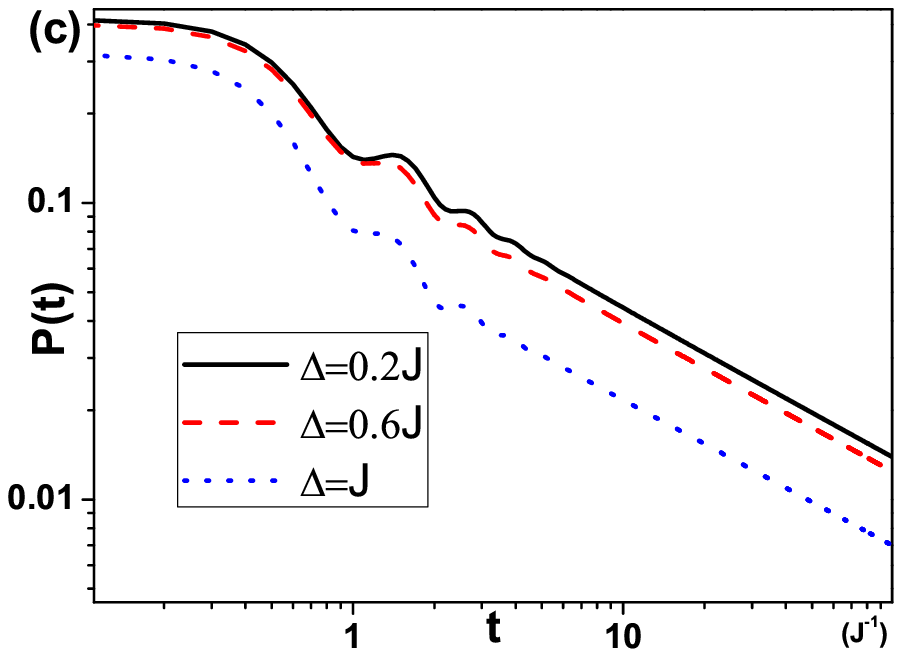}
\includegraphics[width=0.325\linewidth,bb=18 19 278 218]{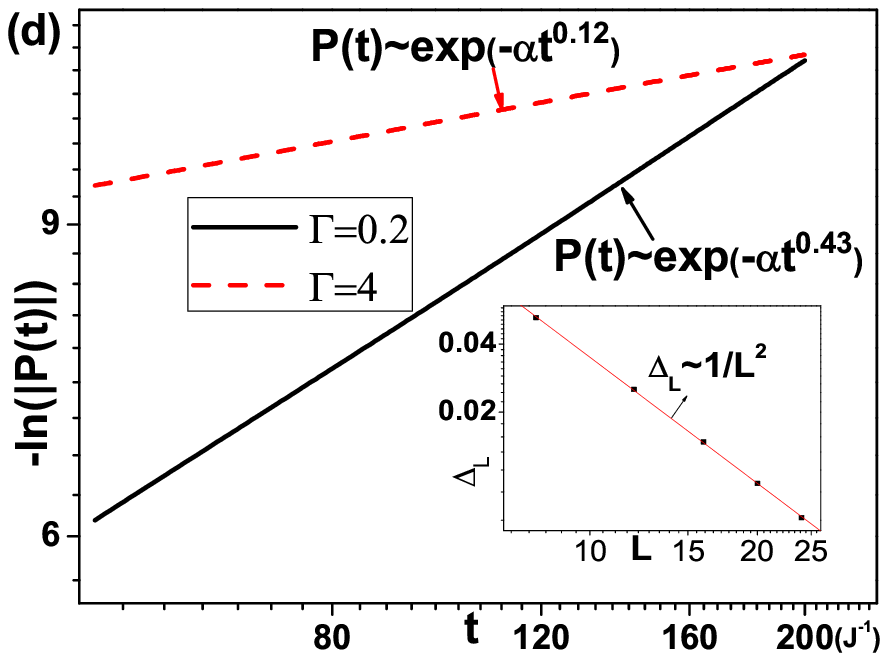}
\includegraphics[width=0.325\linewidth,bb=17 24 276 211]{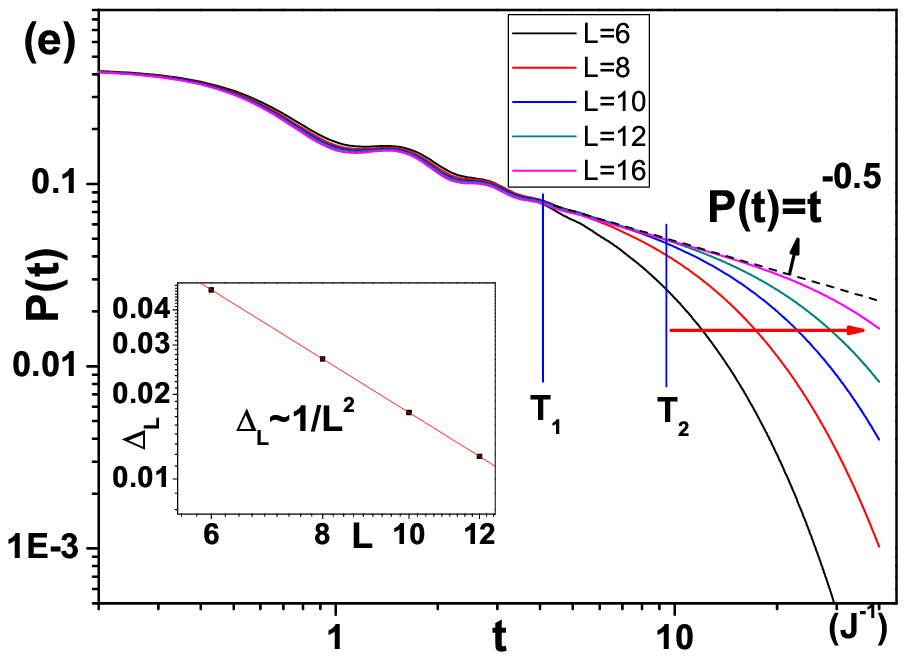}
\includegraphics[width=0.335\linewidth,bb=16 23 280 210]{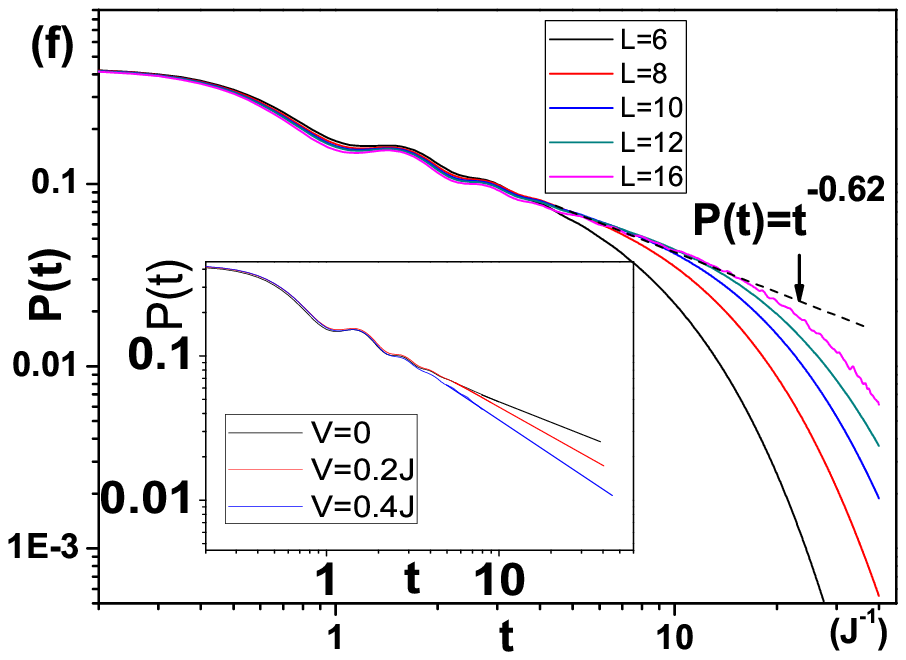}
\caption{(a)The spectrum of the (noninteracting) Liouville superoperator $\mathbb{L}_k$  with parameters $\delta=J$, $\kappa=J$; (b) time evolution of $P(t)$ for the non-interacting case with $\delta=2J$ and different $\kappa$ (the inset is the dynamics of both the diagonal and off-diagonal correlation functions); Dynamics of $P(t)$ in the presence of (c) off-site pairing perturbations and (d) static disorder (the inset is the finite size scaling of the Liouville gap $\Delta_L$ for the noninteracting case, which is obtained by fitting the long-time behavior using the exponential function $P(t)\sim e^{-\Delta_L t}$).   Finite-size effect on the dynamics of $P(t)$ for (e) the noninteracting case  (the inset  is the finite size scaling of $\Delta_L$ in the presence of strong interaction $V=2J$) and (f) the weak interacting case (V=0.2J) (the inset is the extrapolated results for the dynamics in the thermodynamics limit with weak interactions). In both cases open boundary conditions are chosen and the dashed lines are the extrapolated results for the dynamics in thermodynamic limit. For (c)-(f), we choose the parameters $\delta=2J$, $\kappa=2J$.}
\label{fig:fermion}
\end{figure*}

\section{One dimensional  spinless fermion}
\label{sec:1Dfermion}
The first model we consider is a 1D spinless fermonic model with a stochastically fluctuating staggered  potential with the Hamiltonian $H=H_0+H'$, where $H_0$ is the Hamiltonian of the noninteracting fermions with stochastic driving:
\begin{eqnarray}
H_0&=&\sum_{i,\sigma}  -J(c_i^\dag c_{i+1}+\text{h.c.})- \sum_i\lambda(t)(-1)^i n_i .\label{eq:fermion1}
\end{eqnarray}
$c_i^\dag (c_i)$ is the creation (annihilation) operator of the spinless fermion and $n_i=c^\dag_i c_i$ is the local density operator.  $\lambda(t)$ stochastically jumps between $\lambda_a=\delta$ and $\lambda_b=-\delta$ with jumping rate $\kappa$. $H'$ represents different kinds of perturbations that will be explicitly analyzed below.  We assume that initially the system is in the ground state of $H_a$, and focus on the population imbalance $P(t)=\sum_i (-1)^i n_i/L$ ($L$ is the number of lattice sites). As we analyzed above, the infinite-$T$ state always plays the role of an attractor during the time evolution, and we will explore the relaxation dynamics towards this fixed point.

We first consider the unperturbed case ($H'=0$), where the translationally invariant system without interaction is best
treated in a Fourier transformed picture as a collection of independent momentum ($k$) modes ($-\pi/2<k<\pi/2$; we take the lattice constant to be 1).  In the long time limit, each $k$-mode will decay exponentially with time $\sim e^{-\Delta_k t}$, where $\Delta_k$ is the gap of $\mathbb{L}_k$ (the absolute value of the second largest real part of the eigenvalues of $\mathbb{L}_k$),  which vanishes at $k_c=\pi/2$, and can be expanded around $k_c$ as  $\Delta_k=\alpha(\delta k)^2$ with $\delta k=k-k_c$ (the first order of $\delta k$ vanishes for  symmetry reason), as shown in Fig.~\ref{fig:fermion}(a).  The dynamics of physical observables results from the collective behavior of all $k$-modes, and the long-time asymptotic behavior is determined by the long-lived modes ($k$-modes near the gapless point):
\begin{equation}
P(t)\sim \int_0^\infty d\,\delta k \quad e^{-\delta k^2 t}\sim t^{-1/2}. \label{eq:integral}
\end{equation}
This agrees with our numerical results shown in
Fig.~\ref{fig:fermion}(b), that the long-time  behavior of $P(t)$ always decays algebraically with time: $P(t)\sim t^{-\eta}$, with a universal exponent $\eta=1/2$ independent of the system parameters.  As previously analyzed, this universality is related to the absence of a gap in the spectrum of $\mathbb{L}$. In the following, we will consider various perturbations $H'$ including (a) a pairing term breaking the total particle number conservation; (b) static disorder; and (c) nearest neighboring interactions, and study whether they can qualitatively change the universality of the  long-time relaxation dynamics.

{\it Pairing term breaking the total particle number conservation:} A key feature of the above case is that the total particle number is conserved during the time evolution, which corresponds to a continuous symmetry (U(1)) in the Hamiltonian.  It is well-known that in classical critical systems, conservation laws play an important role in determining the universality class of the dynamical critical phenomena: the relaxation behavior, e.g. the dynamic critical exponent, of a system with a non-conserved order parameter (model A according to the convention of HH, e.g. the dynamic Ising model\cite{Glauber1963}) can be significantly different from that of one that obeys the conservation law (HH's model B, e.g. the driven diffusive lattice gases\cite{Schmittmann1995}). For the quantum system we studied above, it is natural to ask whether the relaxation behavior is related to the conservation law of the total particle number. To address this question, we add a perturbing pairing term which breaks particle number conservation, $H'=\Delta\sum_i (c_i^\dag c_{i+1}^\dag+h.c)$. From Fig.~\ref{fig:fermion}(c), we find that contrary to classical critical dynamics, introducing the conservation-law breaking perturbation doesn't qualitatively change the long-time relaxation dynamics, which is still algebraic with $P(t)\sim t^{-1/2}$. Mathematically, this is due to the fact that the paring perturbation doesn't change the analytic properties of $\Delta_k$ near the gapless point.

{\it Effect of static disorder:} Now we study the effect of static disorder on the long-time behavior of this stochastically driven system; $H'=\sum_i V_i n_i$ where $V_i$ represents static disorder
sampled from a uniform random distribution with $V_i\in [-\Gamma,\Gamma]$.  From Fig.~\ref{fig:fermion}(d), we find that even weak disorder ($\Gamma\ll J$) will qualitatively change the long-time behavior of the system from an algebraic decay to a stretched  exponential decay $P(t) \sim \exp (-\alpha t^\beta)$, where $0<\beta<1$ is a non-universal parameter which depends on the parameters in the Hamiltonian. These unconventional relaxation dynamics was first discovered in 1847 by Kohlrausch, and have been observed in various systems such as molecular\cite{Philips1996} and spin\cite{Chamberlin1984} glasses and dissipative interacting quantum systems\cite{Poletti2013,Carmele2015,Levi2015,Gopalakrishnan2016,Fischer2016}. Up to now, considerable theoretical effort has been devoted to understand the origin of the stretched exponential decay\cite{Ngai1979,Palmer1984,Berthier2011}. For the system considered above, we can propose a simple understanding of the unconventional relaxation dynamics by considering an extreme situation, where $V_i$  can only take two discrete values 0 and $\infty$ with the  probability $p$ and $1-p$ ($0<p<1$). In this case, the system turns to a site-dilute model composed of open chains (clusters)  with different lengths. Notice that for a cluster with length $l$, its relaxation time $\tau_l\sim l^2$ (as shown in the inset of Fig.\ref{fig:fermion} d), therefore the long clusters dominate the long-time dynamics of the system. On the other hand, in a site-dilute chain, the appearance of the long clusters is a rare event with a probability exponentially decaying with length $l$: $W_l=p^l=e^{(\ln p) l}$. Under these approximations, the long-time behavior of the system can be obtained as $
P(t)\sim\int dl\, W_l\, l\, e^{-t/\tau_l}=\int dl\, l\, e^{- (c' l+\frac {ct}{l^2})}$
where $c'=-\ln p$. For large $t$, this integral can be evaluated in the saddle-point approximation, and we obtain $P(t)\sim \exp[-\tilde{c}t^{\frac 13}]$.

{\it Effect of interaction:} In all the cases studied previously, the Hamiltonians are of quadratic form. Hence, the information on the system can be obtained from the single-particle correlation functions. To investigate the dynamics of a genuine interacting quantum many-body system,  we consider the perturbation $H'=\sum_i V n_i n_{i+1}$ representing nearest-neighbor (NN) interactions between the spinless fermions. As pointed out previously, for this interacting case, the dimension of the EOM in Eq.~(\ref{eq:Master1}) grows fast with the system size $2\boldsymbol{N}^2\sim \mathcal{O}(e^{2L})$. This makes it  impractical for large systems.  An alternative method is to calculate the unitary evolution for each given stochastic trajectory and then explicitly perform the ensemble average over a sufficiently large number of trajectories. The dimension of the EOM in the unitary evolution method (UEM) $\boldsymbol{N}\sim \mathcal{O}(e^{L})$, even though  much smaller than that in the marginal density matrix method (MDMM), is still exponential in lattice site. To extrapolate the long-time behavior of an interacting quantum system in the thermodynamic limit based on the finite-size results, we need to carefully study the role of finite-size effects.

To gain some insight into the finite-size effects, we first focus on the non-interacting case. From Fig.~\ref{fig:fermion}(e) we find that for a finite-size system the time evolution can be divided into three regimes by two time scales $T_1$ and $T_2$: the short-term dynamics ($t<T_1$) is characterized by the coherence oscillations and depends on the initial state; once the initial state information is lost, the systems enter the intermediate regime ($T_1<t<T_2$) exhibiting a power law behavior. Since any finite system has a nonzero Liouville gap, the finite size effect will dominate the long-term evolution ($t>T_2$), and lead to an exponential decay with time. From Fig.~\ref{fig:fermion}(e) we find that the time scale $T_1$ is insensitive to the system size $L$, while $T_2$ monotonously increases with $L$. Therefore, we expect that in the thermodynamic limit $L\rightarrow \infty$, the long-term exponential dynamics will give way to the intermediate algebraic dynamics, which represents the long-time behavior of the system. This tendency can be seen clearly even for small systems ($L\leq 16$).

For interacting systems, we expect that the above dynamical structure of the time evolution still holds at least for weak interactions $V\ll J$, which allows us to extract the long-time behavior from the intermediate dynamics of finite-size systems. The dynamics of $P(t)$ in the presence of a weak interaction is shown in Fig.~\ref{fig:fermion}(f), where we find that the structure of the dynamic behavior is similar to that of the non-interacting case. However, the exponent of the power-law decay in the intermediate region is changed by the interaction to $P(t)\sim t^{-\beta}$ with $\beta>0.5$, which reminds us of the algebraic correlation functions in the Luttinger liquid whose power-law exponents are also renormalized by interaction\cite{Giamarchi2003}.  The finite size scaling indicates that, similar to the noninteracting case, this intermediate algebraic regime with a renormalized exponent can also be extrapolated to infinite time in the thermodynamic limit. A physical picture is that the interaction makes the momentum modes no longer independent of each other, and the scattering between them leads to transitions between the fast and slow modes, thus makes the decay faster than that in the non-interacting case. Recently, a similar dynamical behavior has been observed in the relaxation dynamics of many-body localized systems\cite{Serbyn2014,Agarwal2015}.  For strong interactions, the dynamical structure is complex and does not resemble the non-interacting case, thus to extract the long-time behavior requires larger systems  which is beyond the capability of the current method. The only information we can obtain about the strong interacting case is that the finite size scaling (inset of
Fig.~\ref{fig:fermion}(e)) indicates that the Liouville gap vanishes in the thermodynamic limit as $\Delta_L\propto 1/L^2$, which precludes the possibility of exponential decay for $L\rightarrow \infty$ in the long-time behavior.

{\it Discussion:} At the end of this section, we add some remarks. Firstly, we shall compare our results of the stochastically driven systems with those in their periodical counterparts, where the periodic driving may either drive the system to time-periodic regimes synchronous\cite{Russomanno2012} or asynchronous\cite{Chandran2015} with the driving, or heat the system to an infinite-temperature state\cite{Alessio2014} after an extraordinary long time, and the asymptotic dynamics depend on lots of details of the systems\cite{Russomanno2012,Alessio2014,Lazarides2014a,Chandran2015,Citro2015}. For the stochastic cases, the external driving will quickly destroy the initial state information and drive the system into the  \textquotedblleft long time\textquotedblright asymptotic regime after a relatively short time $T_1\sim\mathcal{O}(J^{-1})$. Due to the ensemble average, the stochastic driving facilitates the stabilization of the system into an  universal dynamical regime, which enables us to study the dynamical universality. Secondly, since stochastic driving causes decoherence, one might expect that the off-diagonal terms of the density matrix vanish after long-time evolution, and all the previously-studied dynamics near the steady state could be reduced to rate equations of the diagonal matrix elements. Thus, the dynamics would be essentially classic. To clarify this point, we calculate the dynamics of one of the off-site correlations $C_2(t)=\langle c^\dag_ic_{i+2} \rangle$ as a representative of the off-diagonal elements of the density matrix, and compare it to that of the typical diagonal one $P(t)$. As shown in the inset of
Fig.~\ref{fig:fermion}(b), $C_2(t)\sim t^{-0.5}$ decays as slowly as $P(t)$, which indicates that even near the infinite temperature state, the quantum fluctuations still play an important role and the dynamics are not classical.  Finally, it is natural to ask for stochastically driven Hamiltonians with locally bounded Hilbert spaces, whether it is possible to find a long-time behavior different from the scenarios studied above. To address this question, we propose a spinful fermion model (see the Appendix), which exhibits a dynamical phase transition from an algebraic to exponential long-time relaxation behavior by tuning the Hamiltonian parameters. Also, we propose a sufficient condition for the existence of the algebraic relaxation for a general quadratic fermonic system.

\begin{figure*}[htb]
\includegraphics[width=0.325\linewidth,bb=15 25 270 211]{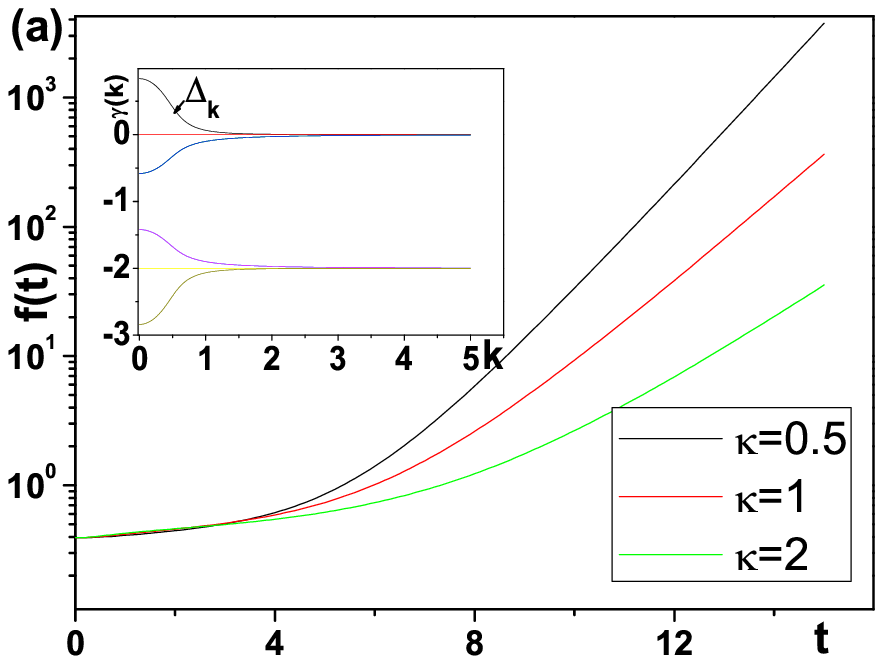}
\includegraphics[width=0.325\linewidth,bb=17 25 270 212]{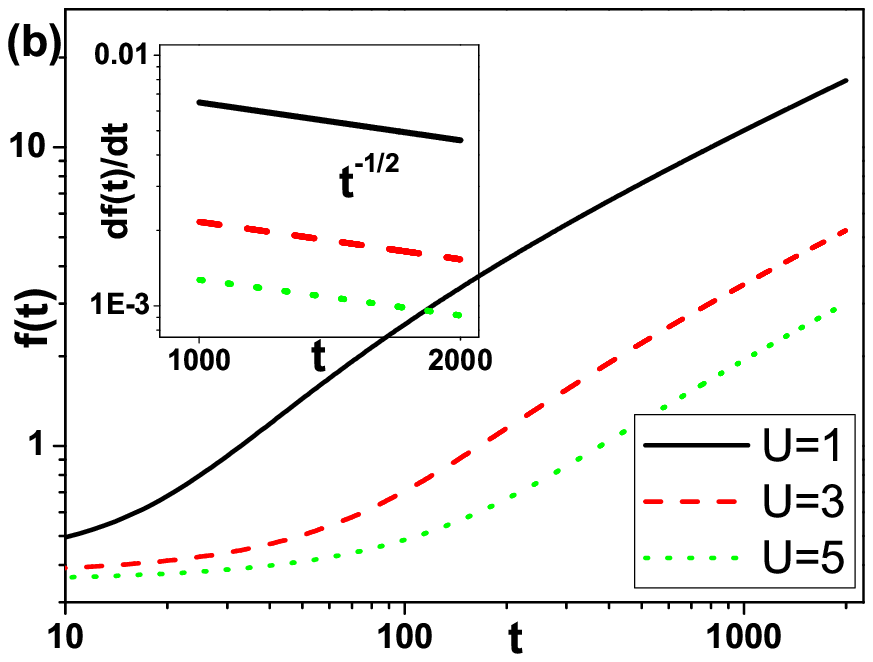}
\includegraphics[width=0.325\linewidth,bb=16 22 276 211]{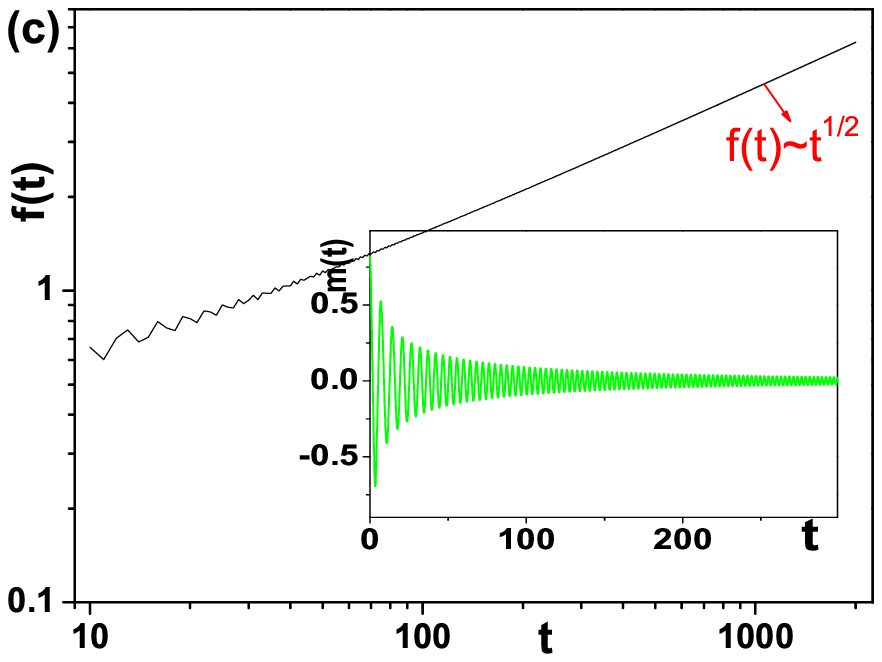}
\caption{(a) Time evolution of $f(t)$ for the non-interacting case with parameters $r_a=-r_b=0.5$ (the inset is the corresponding Liouville spectrum); (b)The dynamics of $f(t)$ starting from the paramagnetic initial state with different $U$ and  $r_a=-r_b=0.5$, $\kappa=1$ (the inset is the long time dynamics of $\dot{f}(t)$) (c) )The dynamics of $f(t)$ starting from the ferromagnetic initial state with  $U=3$, $r_a=-1.5$ and $r_b=-0.5$,  $\kappa=1$ (the inset is the dynamics of the magnetization $m(t)$),  for (a)-(c)  $\Lambda=5$. }
\label{fig:boson}
\end{figure*}

\section{Quantum $O(N)$ model in the large-$N$ limit}
\label{sec:ON}

In the last section, we studied an example with a locally bounded Hilbert space, where the infinite temperature state is well-defined (the unit matrix in the Hilbert space) and all physical observables will converge eventually. However, for systems with a locally unbounded Hilbert space, the stochastic driving force can infinitely heat the systems and the physical observables will diverge with time. Hence, we expect that the long-time behavior will be fundamentally different from the  previously studied cases. As an example, we consider a quantum $O(N)$ model in the large-$N$ limit, which provides a paradigm to understand
symmetry-breaking in statistical mechanics, and both its equilibrium
properties and unitary dynamics can be solved exactly in different dimensions. More precisely, we study a 2D quantum $O(N)$ model with a fluctuating mass that drives the system across the phase boundary of the equilibrium phase diagram. Taking the advantage of infinite $N$, the Hamiltonian of the interacting quantum system can be reduced to a quadratic form with a time-dependent parameter that is self-consistently determined during the time evolution, which allows us to study the dynamics of these genuine interacting quantum many-body systems, {\it e.g}. the quantum quench\cite{Smacchia2015} and periodically driven\cite{Chandran2015} problems.

The Hamiltonian of a quantum $O(N)$ model with a fluctuating mass  reads
\begin{equation}
H=\int d^d\mathbf{x}\frac{|\vec{\pi}(\mathbf{x})|^2}2+\frac{|\nabla \vec{\phi}(\mathbf{x})|^2}2+\frac{r(t)}2|\vec{\phi}(\mathbf{x})|^2+\frac {U}{4N}(|\vec{\phi}(\mathbf{x})|^2)^2\label{eq:Hameff2}
\end{equation}
where $\vec{\phi}(\mathbf{x})=[\phi_1(\mathbf{x}),\cdots \phi_N(\mathbf{x})] $ are $N$-component real vector field operators and $|\vec\phi(\mathbf{x})|^2=\sum_i \phi^2_i(\mathbf{x})$. $\vec{\pi}(\mathbf{x})$ are conjugate field operators of $\vec{\phi}(\mathbf{x})$ that satisfy the commutation relation $[\phi_i(\mathbf{x}),\pi_j(\mathbf{x'})]=\delta_{ij} \delta(\mathbf{x}-\mathbf{x'})$. $r(t)$ is a time-dependent mass term. It randomly jumps between two values $r_a$ and $r_b$ with the transition rate $\kappa$. For $d\geq 2$, the equilibrium critical point between a ferromagnetic  and paramagnetic phases is identified by the condition $r_c=-\frac U4\int_k \frac {1}{k^2} $ (from now on we define  $\int_k=\int^\Lambda \frac{d^d\mathbf{k}}{(2\pi)^d}$ with $\Lambda$ the ultraviolet cutoff in momentum space).

{\it Paramagnetic case:} We first discuss the case where the initial state is prepared the paramagnetic region, where the $O(N)$ symmetry is preserved during the time evolution.  The interaction terms can be decoupled by introducing the auxiliary field $\rho(\mathbf{x},t)$ as
\begin{equation}
 e^{-\frac{U}{4N}(|\vec{\phi}(\mathbf{x})|^2)^2}=\int \mathcal{D}[\rho]e^{-i\frac U2 \rho(\mathbf{x},t)  |\vec{\phi}(\mathbf{x})|^2-\frac {UN}{4} \rho^2(\mathbf{x},t)}.
\end{equation}
For $N\rightarrow \infty$, the fluctuations are suppressed by the large-$N$ effect, and the auxiliary field $\rho(\mathbf{x},t)$ can be replaced by its saddle point value $\rho(\mathbf{x},t)=-if(t)$ with $f(t)=\int d^d\mathbf{x}\langle |\vec{\phi}(\mathbf{x})|^2\rangle/N$.\cite{Moshe2003}
Performing the Fourier transformation $\phi_i(\mathbf{x})=\int_k e^{i\mathbf{k}\cdot\mathbf{x}}\phi_i(\mathbf{k})$ and introducing the ladder operators $a$ and $a^\dag$\cite{Sachdev1999} by $\phi(\mathbf{k})=\frac {1}{\sqrt{2}}(a_{\mathbf{k}}+a_{-\mathbf{k}}^\dag)$, $\pi(\mathbf{k})=\frac {i}{\sqrt{2}}(a_{-\mathbf{k}}^\dag-a_{\mathbf{k}})$, where $[a_\mathbf{k},a_\mathbf{k'}^\dag]=(2\pi)^d \delta(\mathbf{k}-\mathbf{k'})$,  the Hamiltonian turns into
\begin{equation}
H=\int_k  \frac{1+\frac{\delta}2}2 (a^\dag_{\mathbf{k}} a_{\mathbf{k}}+a_{-\mathbf{k}}a_{-\mathbf{k}}^\dag)+\frac \delta 4 (a_\mathbf{k}^\dag a_{-\mathbf{k}}^\dag+a_{-\mathbf{k}}a_{\mathbf{k}}) ,\label{eq:Hampara}
\end{equation}
where $\delta=r(t)+k^2+U f(t)-1$.  In the following, we will focus on the self-consistent field $f(t)=\int_k f_{\mathbf{k}}(t)$ where
$ f_{\mathbf{k}}(t)=\langle \phi(\mathbf{k})\phi(-\mathbf{k})\rangle$ .
To study the time evolution of $f(t)$, we introduce the vector representation of the bosonic correlation functions $\vec{G}_\mathbf{k}=[\langle a_\mathbf{k}^\dag a_\mathbf{k}\rangle, \langle a_\mathbf{k}^\dag a^\dag_\mathbf{-k}\rangle,\langle a_\mathbf{-k} a_\mathbf{k}\rangle, \langle a_\mathbf{-k} a_\mathbf{-k}^\dag\rangle]^T $. As previously analyzed, $\vec{G}_\mathbf{k}=\vec{G}_\mathbf{k}^a+\vec{G}_\mathbf{k}^b$ where $\vec{G}_\mathbf{k}^{a(b)}$ is the correlation functions corresponding to the  marginal density matrix $\rho_{a(b)}$ with the  EOM
\begin{equation}
\frac{d}{dt}\left [
\begin{array}{c}
\vec{G}_\mathbf{k}^a\\
\vec{G}_\mathbf{k}^b
\end{array} \right ]=\mathbb{L}_\mathbf{k}[f(t)]\left [
\begin{array}{c}
\vec{G}_\mathbf{k}^a\\
\vec{G}_\mathbf{k}^b
\end{array} \right ]=
\left [
\begin{array}{cc}
\hat{\Gamma}_{\mathbf{k}}^a-\kappa \hat{\mathbf{1}}  &  \kappa \hat{\mathbf{1}} \\
\kappa \hat{\mathbf{1}} & \hat{\Gamma}_{\mathbf{k}}^b-\kappa \hat{\mathbf{1}}
\end{array}
 \right ]\left [
\begin{array}{c}
\vec{G}_{\mathbf{k}}^a \\
\vec{G}_{\mathbf{k}}^b
\end{array} \right ] \label{eq:EOMpara}
\end{equation}
where $\hat{\mathbf{1}}$ is the unit matrix with dimension 4 and
\begin{equation}
\hat{\Gamma}_\mathbf{k}^{a(b)}=
\left [
\begin{array}{cccc}
0  & -\frac i2 \delta^{a(b)}_\mathbf{k} & \frac i2 \delta^{a(b)}_\mathbf{k} & 0 \\
\frac i2 \delta^{a(b)}_\mathbf{k} & i(2+\delta_\mathbf{k}^{a(b)}) & 0 &\frac i2 \delta^{a(b)}_\mathbf{k}\\
 -\frac i2 \delta^{a(b)}_\mathbf{k} & 0 & -i(2+\delta_\mathbf{k}^{a(b)}) &  -\frac i2 \delta^{a(b)}_\mathbf{k}\\
 0  & -\frac i2 \delta^{a(b)}_\mathbf{k} & \frac i2 \delta^{a(b)}_\mathbf{k} & 0
\end{array}
 \right ]
\end{equation}
in which $\delta^{a(b)}_\mathbf{k}=r_{a(b)}+k^2+U f(t)-1$  can be determined self-consistently during the time evolution.

{\it Ferromagnetic case:} If we start from the ground state in the ferromagnetic region, the $O(N)$ symmetry has already spontaneously been broken from the beginning. Without loss of generality, we assume that symmetry is broken along the 1-direction in
order parameter space,  thus $\phi_1(\mathbf{x})$ contains a finite uniform magnetization $m(t)=\langle \phi_1(\mathbf{x})\rangle /\sqrt{N}$, and the field $\phi(\mathbf{k})$ can be expressed in terms of ladder operators as   $\phi_i(\mathbf{k})=\frac {1}{\sqrt{2}}(a_{\mathbf{k}}+a_{-\mathbf{k}}^\dag)+\delta_{i1}\phi_0(t)$ with $\phi_0(t)=\sqrt{N} m(t)$\cite{Sachdev1999}. The corresponding self-consistent Hamiltonian takes the same form as that of the paramagnetic case Eq.~(\ref{eq:Hampara}), with the only difference that the $\delta(t)$ in Eq.~(\ref{eq:Hampara}) is replaced by $\tilde{\delta}(t)=r(t)+k^2+ U(\tilde{f}(t)+m^2(t))-1$, where $\tilde{f}(t)=\int d^d\mathbf{x} \sum_{i=2}^N \frac 1N  \langle\phi_i^2(\mathbf{x})\rangle.$
The EOM of the correlation functions are similar to the paramagnetic case Eq.~(\ref{eq:EOMpara}), with $\delta^{a(b)}_\mathbf{k}$ replaced by $\tilde{\delta}_\mathbf{k}^{a(b)}=r_{a(b)}+k^2+ U(\tilde{f}(t)+m^2(t))-1$, and the EOM for the magnetization  $m(t)$ can be obtained from that for $\phi_0^{a(b)}$:
\begin{equation}
\frac{d}{dt}\left [
\begin{array}{c}
\langle \phi_0^a\rangle\\
\langle \pi_0^a\rangle\\
\langle \phi_0^b\rangle\\
\langle \pi_0^b\rangle
\end{array} \right ]=
\left [
\begin{array}{cccc}
-\kappa  &  1 & \kappa &0\\
-\tilde{\delta}^a_0 & -\kappa &0 &\kappa\\
\kappa  &  0 & -\kappa & 1\\
0 & \kappa & -\tilde{\delta}^b_0 &-\kappa
\end{array}
 \right ]\left [
\begin{array}{c}
\langle \phi_0^a\rangle\\
\langle \pi_0^a\rangle\\
\langle \phi_0^b\rangle\\
\langle \pi_0^b\rangle
\end{array} \right ]
\end{equation}
where $\sqrt{N} m(t)=\langle\phi_0\rangle=\langle\phi_0^a\rangle+\langle\phi^b_0\rangle$.

{\it Results:} We first focus on the noninteracting case ($U=0$), as shown in Fig.~\ref{fig:boson}(a). This bosonic system with a locally unbounded Hilbert space will absorb energy indefinitely, which leads to exponentially divergent dynamics due to the parametric resonance between the external driving and the selected momentum modes of the Hamiltonian.  Mathematically, the exponential divergence indicates a positive branch in the spectrum of the Liouville superoperator $\mathbb{L}_k$ defined in Eq.~(\ref{eq:EOMpara}), as shown in the inset of Fig.~\ref{fig:boson}(a).
In the presence of interaction ($U>0$), we find that even though the dynamics are still divergent with time, the interaction will fundamentally change the divergence from an exponential to an algebraic one in the long time dynamics: $f(t)\sim t^\eta$, where again the exponent $\eta=0.5$ is universal and independent of the details of the systems, e.g.\ the parameters in the Hamiltonian and external driving, the strength of the interaction as well as the choices of the initially state, as shown in Fig.~\ref{fig:boson}(b) and (c). Physically, this means that in the quantum $O(N)$ model in the large-$N$ limit, the (repulsive) interaction will significantly suppress the driving-induced heating dynamics through a nonlinear effect: the divergence of $f(t)$ will increase the effective mass of the system which, on the other way, makes the system less and less sensitive to the external driving. Mathematically,  the EOM of each k-mode  is determined by $\Delta_\mathbf{k}$:
\begin{equation}
df_\mathbf{k}(t)/dt=\Delta_\mathbf{k}[f(t)] f_{\mathbf{k}}(t)
\end{equation}
where the gap $\Delta_\mathbf{k}$ is the real part of the positive eigenvalue of the instantaneous Liouville superoperator $\mathbb{L}_\mathbf{k}[f(t)]$ defined in Eq.~(\ref{eq:EOMpara}). Different modes are coupled through the relation $f(t)=\int_k f_{\mathbf{k}}(t)$. Since $f(t)$ diverges with time, in the long time limit we have $Uf(t)\gg r_{a(b)}+\Lambda^2$ and $\delta^{a(b)}_\mathbf{k}\approx U f(t)\pm \delta r$, where $\delta r=(r_a-r_b)/2\ll Uf(t)$. Therefore, based on the perturbation analysis, the gap can be approximated as $\Delta_\mathbf{k}[f(t)]\sim 1/(Uf(t))^2$.  Hence, we can obtain the EOM of $f(t)$ in the long time limit as
\begin{equation}
df(t)/dt\propto 1/(U^2f(t))
\end{equation}
with the asymptotic solution $f(t)\sim t^{\frac 12}$ for $t\rightarrow \infty$. For the ferromagnetic case, we can find that in the long-time limit, the spontaneous magnetization is destroyed by external stochastic driving; $m(t)\rightarrow 0$ as shown in the inset of Fig.~\ref{fig:boson}(c). Therefore, the long-time behavior if we start from a ferromagnetic initial state is qualitatively the same as if starting from the paramagnetic case.

\section{Experimental realization}
\label{sec:experiment}
In this section, we will briefly discuss the possible experimental realization of the above two models.   The 1D spinless fermonic model with a stochastically fluctuating staggered potential can be realized by loading ultracold fermions (or hard-core bosons) into quasi-1D optical superlattice potential, which can be implemented by overlaying two commensurate lattices generated by lasers at the wavelengths of $\lambda$ and $2\lambda$. The telegraph-like stochastic driving can be artificially introduced in a controlled way by programmable tuning of the relative strength of the two laser beams during the time evolution. In Sec.\ref{sec:1Dfermion}, we focus on the population imbalance between the two sublattices, which can be measured directly by employing a band-mapping and imaging technique\cite{Trotzky2012,Schreiber2015}. The three different perturbations we considered in Sec.~\ref{sec:1Dfermion} can be implemented as follows: the off-site pairing term is a key ingredient to realize the Kitaev model in cold atom systems, and can be implemented by employing a Raman induced dissociation technique and immersing the system into an atomic BCS reservoir formed by Feshbach molecules.\cite{Nascimbene2013,Kraus2012,Hu2014} The static disorder can be created optically by using speckle patterns.\cite{Billy2008}  The NN interactions naturally exist in magnetic dipolar atomic systems in optical lattices. In a recent experiments with Erbium atoms\cite{Baier2015}, it was measured that the strength of the NN interaction ($V/\hbar\simeq 30$ Hz for a lattice constant $a_0=266$nm) is of the same order of magnitude as that of the single-particle hopping amplitude ($J/\hbar \simeq 30\sim 100$ Hz depending on the lattice depths). The parameters  encountered in the experiment typically meet the parameter regime studied previously.

Finally, we will briefly discuss the experimental imperfection conditions and their effect on the long-time behavior. The most common perturbation in the optical lattice setup is the noise, which is another stochastic process independent of the driving protocols and inevitably due to the impurity of the laser beams. In the Appendix, we show that the white noise will not qualitatively change the long-time behavior of the stochastically-driven model. Another common imperfection are the effects of finite temperature. Even though temperature is not well-defined during the non-equilibrium dynamics, it can indeed affect the preparation of the initial state. However, throughout this paper, we focus on the long-time behavior of the system, where the initial state information has been washed out by the external driving;  thus finite-temperature effects are also irrelevant for our results.

The connection between the discussions in Sec.~\ref{sec:ON} and realistic experimental systems is subtle, since any realistic systems has an upper bound of the locally Hilbert space. However, for those systems with a sufficient large local Hilbert space, e.g.\ the multicomponent Bose-Hubbard model with a large component number and high filling factors, we conjecture that the long-time behavior discussed in Sec.~\ref{sec:ON} can capture the correct intermediate-time dynamics during which the information of the initial state has been lost but the energy of the system is still far from its upper bound, because during this time period the system can absorb energy  \textquotedblleft infinitely\textquotedblright\ without feeling the restriction imposed by the upper bound of the local Hilbert space.

\section{Conclusion and Outlook}
\label{sec:conclusion}

In this paper, we study the long-time behavior of stochastically driven quantum many-body systems based on two specific examples. As a conclusion of this paper, we wish to emphasize some connections and differences of our results with other relevant ones and provide an outlook.
First, even though the divergence of the relaxation rate resembles the dynamical critical phenomena in classical systems\cite{Hohenberg1977}, there are two significant differences: (a) the divergences in classical system only occur at the critical point, while in our case the algebraic relaxation holds for a whole parameter regime, (b) conservation laws, which play a key role in determining the universality of the dynamical critical phenomena, seem irrelevant for the relaxation dynamics in our case. Moreover, the stochastically driven systems studied above differ from another well-studied problem, quantum many-body systems subject to white noise, in two aspects: the external driving is spatially homogeneous instead of site-dependent, and the correlation time of the stochastic force is finite ($1/\kappa$) rather than zero (thus is colorful noise). These differences give rise to significant consequences: e.g. it is known that for a 1D XXZ model subject to white noise, the $U(1)$ symmetry breaking term is a relevant perturbation for the long-time behavior while the NN interaction is not\cite{Cai2013}, which is exactly the opposite of our observation in the stochastic driving cases.

By considering specific examples,  we have taken the first step towards characterizing the long-time dynamics of stochastically driven interacting quantum systems, but a comprehensive understanding of this problem is far from achieved. Some avenues for further work immediately suggest themselves. The first and most important question is the generality of the above results derived from specific examples; to what extent can they be applied to other systems with different Hamiltonians, driving protocols, or perturbations? A systematic answer to this question requires us to treat the infinite temperature-state as a fixed point and develop an effective non-equilibrium field theory to characterize the dynamics towards the fixed point, and determine the relevancy of various perturbations through the renormalization group analysis, which is beyond the scope of this work and will be left for the future. From the numerical point of view, to approach the long time behavior for a large system, it would be important to develop efficient numerical methods based e.g.\ on the density matrix renormalization group (DMRG) technique\cite{Schollwock2005,Daley2005} to directly solve the master equation~(\ref{eq:Master1}) instead of doing the ensemble average over all the stochastic trajectories.

{\it Acknowledgements -- } Z.C. wish to thank P. Zoller for  raising  our  interest  to  the problem of stochastically driven quantum many-body systems  and  for  many  stimulating  discussions and valuable suggestions during the work.  We wish to thank  M. A. Baranov and Ying Hu for fruitful discussions.  This work is supported by Austrian Science Fund through SFB FOQUS (FWF Project No. F4006-N16) and the ERC Synergy Grant UQUAM. Z. C. also acknowledges the support from the startup funding in Shanghai Jiao Tong University as well as the NSF of China under Grant No.11674221. C. H. acknowledges support from the Nanosystems  Initiative Munich(NIM)and the ExQM graduate school of the Elitenetzwerk Bayern.

\appendix

\begin{figure*}[htb]
\includegraphics[width=0.325\linewidth,bb=15 25 290 211]{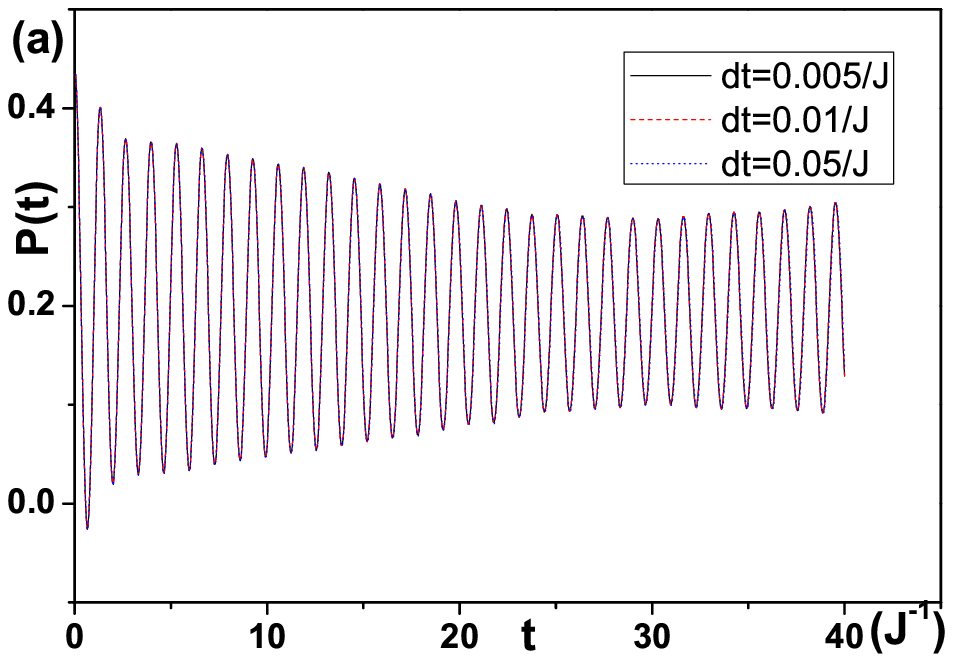}
\includegraphics[width=0.325\linewidth,bb=15 25 290 211]{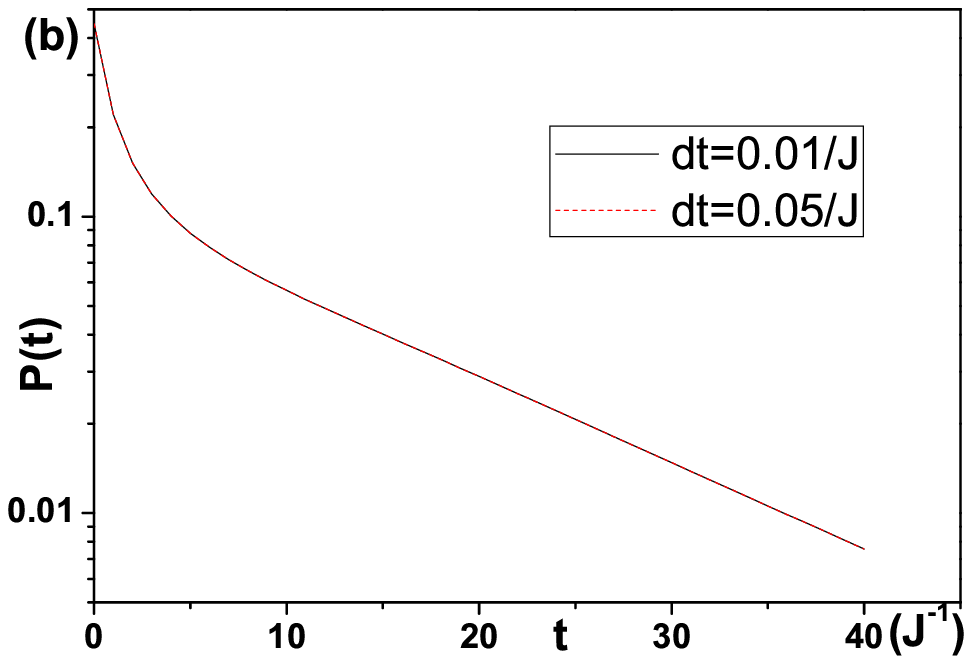}
\includegraphics[width=0.325\linewidth,bb=15 25 290 211]{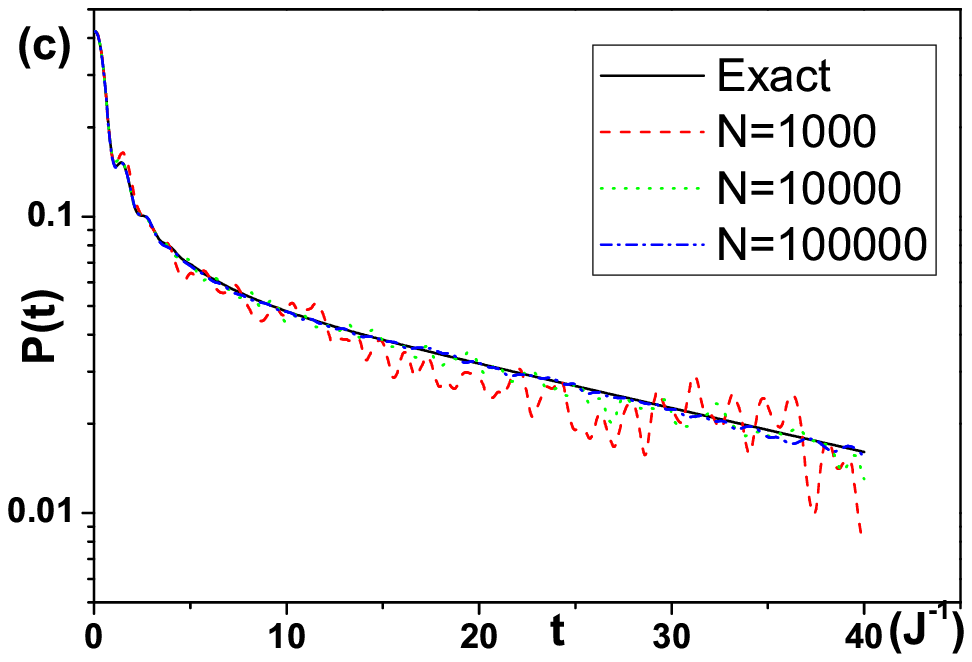}
\includegraphics[width=0.325\linewidth,bb=15 25 290 211]{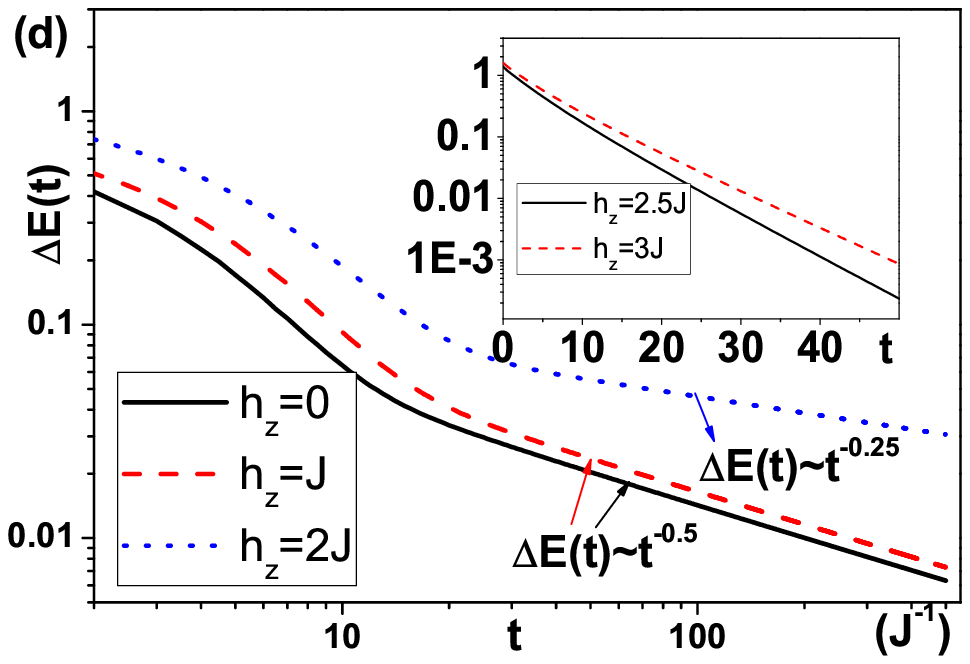}
\includegraphics[width=0.325\linewidth,bb=15 25 290 211]{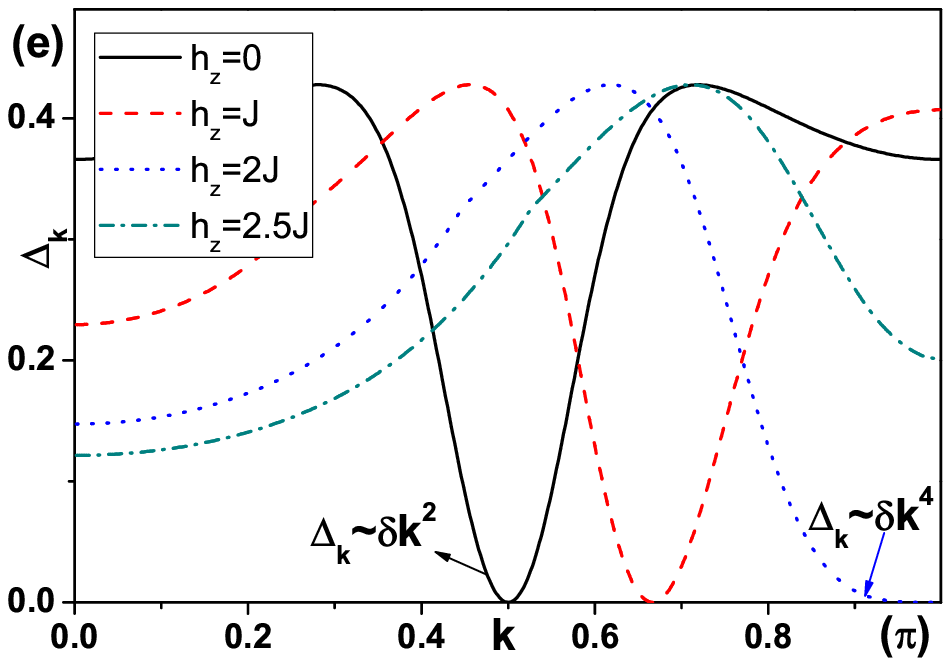}
\includegraphics[width=0.325\linewidth,bb=15 25 290 211]{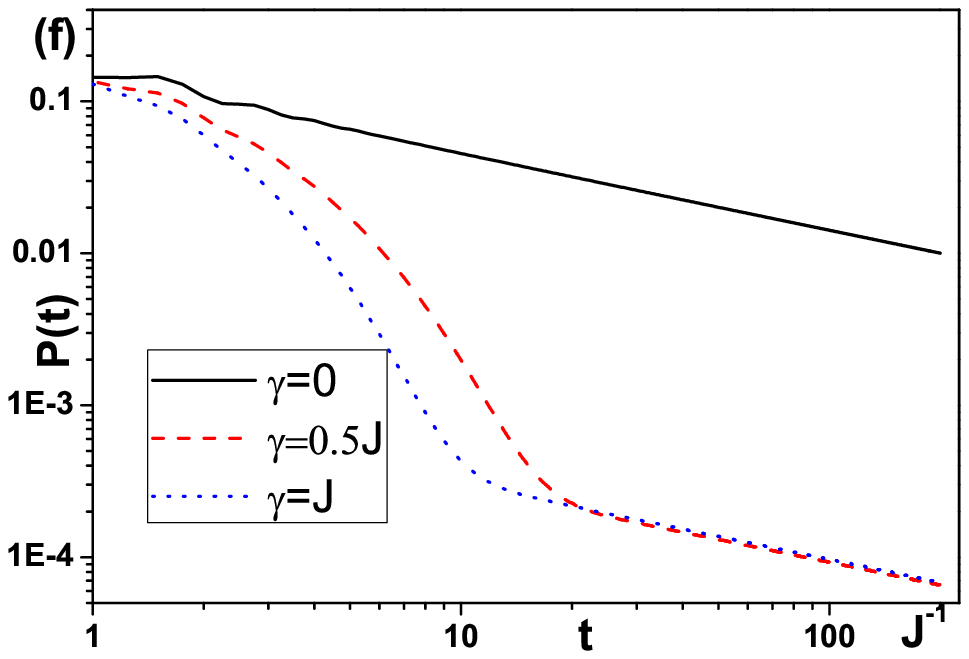}
\caption{Convergence check of the dependence of results on (a) the time step $\delta t$ in UEM with $V=J$, $L=16$; (b) $\delta t$ in the MDMM with $V=0.2J$, $L=12$ and (c) the number of the sampled trajectories $N$ in UEM with $V=0$, $L=16$. In (a)-(c) we choose $\delta=2J$, $\kappa=2J$.  (d) Dynamics of the stochastically driven spin-1/2 fermionic model (Eq.~\ref{eq:spinful}) in algebraic and  exponential (the inset) regions; (e) the Liouville gap $\Delta_k$ for different $h_z$. In (d)-(f), we choose $\lambda_1=0$, $\lambda_2=J$ and $\kappa=2J$. (f) Dynamics of the stochastically driven spinless fermionic model in the presence of external white noise.}
\label{fig:Supp}
\end{figure*}
\section{Derivation of the EOM of the marginal density matrix}

In this section we will derive the EOM of the marginal density matrix (Eq.~(1) in the manuscript). To do that, we first discretize the time axis  (from $t_0$ to $t_n$) into small slices of size $dt=(t_n-t_0)/n$, and we denote $t_k=t_0+k dt$ with k an integer from 0 to $n$. For $n\rightarrow \infty$, we can assume that the jumping of the parameter $\lambda$ can only take place at the discrete time $t_k$. We further denote $\{\lambda_n\}=\{\lambda_1\cdots \lambda_{n-1}\}$ as a trajectory of the fluctuating parameter $\lambda(t)$, which satisfies $\lambda(t_k)=\lambda_k$ for $k=0,\cdots,n-1$ and $\lambda_k=\lambda_a$ or $\lambda_b$. For a given trajectory $\{\lambda_n\}$, we can define  its probability $P_{\{\lambda_n\}}=P(\lambda(t_{n-1})=\lambda_{n-1},t_{n-1};\cdots;\lambda(t_0)=\lambda_0,t_0)$, and the density matrix at $t_n$ following this trajectory as $\rho_{\{\lambda_n\}}(t_n)=\mathcal{U}_{\{\lambda_n\}} \rho(t_0)\mathcal{U}^{-1}_{\{\lambda_n\}}$, where the corresponding unitary evolution operators: $\mathcal{U}_{\{\lambda_n\}}=e^{idtH_{\lambda_{n-1}}}\cdots e^{idtH_{\lambda_0}}$. With these definitions, we obtain the density matrix after the ensemble average of all the trajectories:
\begin{equation}
\rho_s(t_n)=\sum_{\{\lambda_n\}} P_{\{\lambda_n\}} \rho_{\{\lambda_n\}}(t_n) .
\end{equation}
The corresponding marginal density matrix $\rho_{a(b)}(t_n)=\langle \rho(t_n)\delta(\lambda (t_{n-1})=\lambda_{a(b)})\rangle$ can be expressed as
\begin{eqnarray}
\nonumber \rho_a(t_n)=\sum_{\{\lambda_{n-1}\}}P(\lambda_a,t_{n-1}|\lambda_{n-2},t_{n-2})P_{\{\lambda_{n-1}\}}
\\ \times e^{idtH_a}\rho_{\{\lambda_{n-1}\}}(t_{n-1})e^{-idtH_a} ,\label{eq:S2}
\end{eqnarray}
where $P(\lambda_a,t_{n-1}|\lambda_{n-2},t_{n-2})$ denotes the conditional probability of the case that $\lambda(t_{n-1})$ takes the value of $\lambda_a$ if $\lambda(t_{n-2})=\lambda_{n-2}$. To express $\rho_a(t_n)$ in terms of $\rho_a(t_{n-1})$ and $\rho_b(t_{n-1})$, we further expand the summation over $\lambda_{n-2}$ in Eq.~(\ref{eq:S2}) and obtain:
\begin{eqnarray}
\nonumber \rho_a(t_n)=P(\lambda_a,t_{n-1}|\lambda_a,t_{n-2})e^{idt H_a}\rho_a(t_{n-1})e^{-idt H_a}
\\ +P(\lambda_a,t_{n-1}|\lambda_b,t_{n-2})e^{idt H_a}\rho_b(t_{n-1})e^{-idt H_a} .\label{eq:S3}
\end{eqnarray}
We use that in the time interval $[t_n,t_{n-1}]$ the transition probability for the parameter $\lambda(t)$ is $\kappa dt$, which indicates that $P(\lambda_a,t_{n-1}|\lambda_a,t_{n-2})=1-\kappa dt$ and $P(\lambda_a,t_{n-1}|\lambda_b,t_{n-2})=\kappa dt$. In the limit of $dt\rightarrow 0$, we can expand the right-hand side of Eq.~(\ref{eq:S3}) to the first order in $dt$ and obtain:
\begin{equation}
\nonumber \frac{\rho_a(t_n)-\rho_a(t_{n-1})}{dt}= i[\rho_a(t_{n-1}),H_a]-\kappa\rho_a(t_{n-1})+\kappa \rho_b(t_{n-1})
\end{equation}
which reduces to the EOM of the marginal density matrix $\rho_a(t)$,  Eq.~(1) in the manuscript, in the limit $dt\rightarrow 0$ (The EOM of $\rho_b(t)$ can be obtained similarly).

\section{Details of numerical methods}

In Sec.~\ref{sec:1Dfermion}, to study the time evolution of the system, we used two methods: the marginal density matrix method (MDMM)  for small system sizes ($L\leq 12$) and the unitary evolution method (UEM) for larger ones ($L=16$). In this section, we provide some details about these two methods, and check the numerical convergence of the results. We also numerically verify that the MDMM is equivalent to UEM for a sufficiently large number of sampled trajectories.

{\it Unitary evolution method:}  For a given trajectory, the time evolution is unitary under a time-dependent Hamiltonian $H(\{\lambda (t)\}$, thus at time $t$ the wavefunction can be expressed as
\begin{equation}
|\Psi(t)\rangle=\mathcal{T}e^{-i\int_0^t H(\lambda (t'))dt'} |\Psi(0)\rangle=\prod_n e^{-i H_n \delta t}|\Psi(0)\rangle ,
\end{equation}
where $\mathcal{T}$ is the time ordering operator, $\delta t= t/N$ is the time interval and $H_n=H(\lambda (t_n))$ is the Hamiltonian at $t_n$. To calculate the time evolution, it is more convenient to so decompose the total Hamiltonian into pieces that act only on odd bonds and even bonds,
\begin{equation}
H_n=H^n_{even}+H^n_{odd} ,
\end{equation}
where $H^n_{odd}=\sum_i H^n_{2i-1,2i}$ and $H^n_{even}=\sum_i H^n_{2i,2i+1}$. All the terms within the summation of $H_{odd}$ or $H_{even}$ commute with each other. For each time step, the evolution operator can be expanded in a second order Suzuki-Trotter expansion:
\begin{equation}
\nonumber e^{-iH_n \delta t}=e^{-iH^n_{even} \delta t/2}e^{-iH^n_{odd} \delta t}e^{-iH^n_{even} \delta t/2}+\mathcal{O}(\delta t^3) .
\end{equation}
For each bond, we can decompose the Hamiltonian into the diagonal and off-diagonal parts:
\begin{equation}
H^n_{i,i+1}=H^d_{i,i+1}+H^o_{i,i+1}
\end{equation}
where in our case $H^o_{i,i+1}=-J(c_i^\dag c_{i+1}+\text{h.c.})$ and $H^d_{i,i+1}=V n_i n_{i+1}+\frac{\lambda(t_n)} 2 (-1)^i(n_i-n_{i+1})$.  Thus for each bond the evolution operator can be further decomposed as:
\begin{equation}
\nonumber e^{-iH^n_{i,i+1} \delta t}=e^{-iH^d_{i,i+1} \delta t/2}e^{-iH^o_{i,i+1} \delta t}e^{-iH^d_{i,i+1} \delta t/2}+\mathcal{O}(\delta t^3)
\end{equation}
where $e^{-iH^d_{i,i+1} \delta t/2}$ is a diagonal matrix, and $e^{-iH^o_{i,i+1}}$ only operates on two adjacent sites $i$ and $i+1$. Hence, its operation on the wave function can be easily performed without explicitly calculating the matrix $e^{-iH^o_{i,i+1}}$.

In our simulation of the $L=16$ system using UEM, we choose the time interval $\delta t=0.005J^{-1}$. Since the 2nd Trotter decomposition we used gives the errors of 3rd order of the time step $\delta t$, it is necessary to check the numerical convergence of our results in the $\delta t$ we used. To do this, we choose the unitary evolution under one of the simplest trajectory with only one flipping of the parameters at $t=0$ (quantum quench problem). The convergence analysis is shown in Fig.~\ref{fig:Supp}(a).

{\it Marginal density matrix method:} For small system, we can directly solve the EOM of the marginal density matrix Eq.~(\ref{eq:Master2}), which is a linear differential equations, using a 4th order Runge-Kutta method, whose accuracy also depends on the time interval $\delta t$.   In our simulation using MDMM, we choose $\delta t =0.05 J^{-1}$, and the convergence analysis is shown in Fig.~\ref{fig:Supp}(b). Notice that the $\delta t$ we choose in UEM is much smaller than that in the MDMM. The reason is that in the unitary evolution the error introduced by the finite $\delta t$ in the Trotter decomposition will accumulate during the time evolution, and will eventually make the simulation inaccurate, while in MDMM, the evolution is not unitary and it will converge to a steady state. This convergence provides a self-correction mechanism for the accumulated error in the long time evolution, thus allowing us to use larger $\delta t$.

{\it  Equivalence of the two methods:} In the last section, we provide an analytic proof of the equivalence between the above two methods. Here, we will numerically verify that the MDMM is equivalent to UEM for a sufficient large number of sampled trajectories (in our simulation of $L=16$ using UEM, we choose $\mathcal{N}=10^5$). As shown in Fig.~\ref{fig:Supp}(c), the results of UEM will converge to that of MDMM with increasing number of sampled trajectories ($\mathcal{N}$), and for $\mathcal{N}\sim 10^5$, the results of the two methods almost coincide.

\section{Significant others}
In this section, we consider another stochastically driven fermionic model, which exhibits a dynamical phase transition between phases with algebraic and exponential relaxation behavior  absent in the one previously studied in Sec.~\ref{sec:1Dfermion}. Also we propose a sufficient condition for the existence of the algebraic relaxation behavior in general quadratic fermonic systems with stochastic driving.

{\it Dynamical phase transition:} The model we consider is a 1D spin-1/2 fermionic model  with a spin-flip term in a Zeeman field with the Hamiltonian
\begin{eqnarray}
\nonumber H&=&\sum_i [\sum_{\sigma}(-J_\sigma c_{i\sigma}^\dag c_{i+1\sigma}+h.c)\\
&-&h_z (n_{i\uparrow}-n_{i\downarrow})+\lambda(t)(c_{i\uparrow}^\dag c_{i\downarrow}+h.c)]\label{eq:spinful}
\end{eqnarray}
where $\sigma=\uparrow,\downarrow$ denotes the spin index,  $J_\sigma$ is the spin-dependent hopping amplitude, $h_z$ is the Zeeman field and $\lambda(t)$ is the stochastic driving parameter with transition between two values $\lambda_a$ and $\lambda_b$  during the time evolution. Assuming  $J_\uparrow=-J_\downarrow=J$ in the following, we focus on dynamics of the quantity $\Delta E(t)=E(t)-E_0$ to monitor the long-time behavior, where $E(t)= \text{Tr} \rho_s(t) \bar{H}$ with $\bar{H}=(H_a+H_b)/2$ is the time-independent Hamiltonian and $E_0=E(t\rightarrow \infty)$ is the energy at the infinite temperature state (steady state). The long-time behavior of $\Delta E(t)$ characterizes how the system approaches to the steady state.

The dynamics of $\Delta E(t)$ is plotted in Fig.~\ref{fig:Supp}(d) and (e), where we find that for $h_z<2J$, the long-time behavior is similar as before, $\Delta E(t)\sim t^{-\frac 12}$, while at the point $h_z=2J$, it suddenly turns to $\Delta E(t)\sim t^{-\frac 14}$. Beyond this point $h_z>2J$, an exponential decay takes place: $\Delta E(t)\sim e^{-\gamma(h_z) t}$. As previously pointed out, in the absence of interactions, the long-time behavior is determined by the Liouville spectrum $\mathbb{L}_k$. In Fig.~\ref{fig:Supp}(f), we plot the Liouville gap $\Delta_k$, which shows that in the region $0<h_z<2J$, the gapless point is shifted from $k_c=\pi/2$ to $\pi$, while around the gapless point, $\Delta_k$ can always been expanded as $\Delta_k=\alpha \delta k^2+\mathcal{O}( \delta k^4)$ with $\delta k=k-k_c$, which is responsible for the algebraic behavior $\Delta E(t)\sim t^{-\frac 12}$. At the critical point ($h_z=2J$), the coefficient before the quadratic term vanishes ($\alpha=0$) and the quartic term dominates ($\Delta_k=\eta \delta k^4)$. By performing a similar integral as Eq.~(\ref{eq:integral}), we can obtain $\Delta E(t)\sim t^{-\frac 14}$. For $h_z>2J$, a gap is opened indicating an exponential decay.  In summary, we propose a stochastically-driven model exhibiting a dynamical phase transition, which is supplementary to various long-time behaviors in the case previously studied in Sec.~\ref{sec:1Dfermion}.  Since the steady states are the same for both phases,  this phase transition can only be characterized by the dynamical instead of static properties, {\it e.g.} the relaxation time, as well as the singularity of the Liouville gap at the transition point. Notice that the dynamical phase transition we proposed here  has subtle difference from its conventional definition that singularity occurs during the time evolution\cite{Heyl2013,Heyl2015,Zhang2016}.

{\it A sufficient condition for the existence of the algebraic relaxation:}  As previously analyzed, the long-time algebraic relaxation indicates a gapless Liouville superoperator. It is a highly non-trivial question to determine whether a general Liouville superoperator as defined in Eq.~(\ref{eq:Master2}) is gapped or gapless; it has no general answer. However, for a fermionic system with quadratic Hamiltonian, we can propose a sufficient condition for the existence of a gapless Liouville superoperator. We take the translationally invariant system as an example, and the results can be easily generalized to other quadratic Hamiltonians. For a translationally invariant system, the Hamiltonian can be decoupled into independent $k$-mode: $
H=\sum_k \mathbf{C}^\dag_k \hat{H}(k) \mathbf{C}_k$, in which the Hamiltonian of each $k$-mode $\hat{H}_k$ can be decomposed into two parts $\hat{H}(k)=\hat{H}_1(k)+\lambda(t)\hat{H}_2(k)$ with $\hat{H}_1(k)$ corresponding to the non-driven part of the Hamiltonian, and $\lambda(t) \hat{H}_2(k)$ corresponding to the external driving which can be telegraph or other types. If there exists a $k$-mode at $k=k_c$ satisfying $[\hat{H}_1(k_c),\hat{H}_2(k_c)]=0$, then the gap of the Liouville superoperator closes at $k=k_c$. This can be easily understood as following. We start from one of the eigenstates of $\hat{H}_1(k_c)$, which commutes with the external driving Hamiltonian $\hat{H}_2(k_c)$. This indicates that this eigenstate will be immune to the external driving. In other words, a nontrivial steady state other than the infinite-temperature one exists at $k_c$. This degeneracy at $k_c$ and the continuity properties of the Liouville spectrum indicate that the spectrum is gapless. Take Eq.~(\ref{eq:fermion1}) as an example: at $k_c=\pi/2$, $\hat{H}_1(k_c)=0$, which obviously commutes with $\hat{H}_2(k_c)=\lambda(t) \hat{\sigma}_x$, thus the corresponding Liouville gap closes at $k_c=\pi/2$. For systems without translational symmetry, the corresponding Liouville superoperator is gapless for a similar reason as analyzed above, if there exists a single-particle state $|\varphi\rangle$  which is simultaneously an eigenstate of $\hat{H}_1$ and $\hat{H}_2$.

\section{External white noise}

In this section, we analyze the fate of the stochastically-driven spinless fermion model in the presence of white noise, which provides another stochastic process independent of the driving force. The Hamiltonian of the white noise can be written as $H'=\sum_i \xi_i(t) n_i$, where $\xi_i(t)$ represents the site-dependent random field satisfying $\langle \xi_i(t) \xi_j(t')\rangle=\sqrt{\gamma}\delta_{ij}\delta(t-t')$. In spite of the stochastic properties, the white noise differs from the previously-studied stochastic driving in two aspects: it is site-dependent and with zero correlation length. The ensemble average over the external white noise can be performed following the standard procedure\cite{Horstmann2013}, and the EOM of the marginal density matrix reads:
 \begin{eqnarray}
 \nonumber \frac{d\rho_a(t)}{dt}=i[\rho_a,\hat{H}_a]+\gamma \hat{\mathcal{D}}\rho_a-\kappa\rho_a+\kappa\rho_b\\
 \frac{d\rho_b(t)}{dt}=i[\rho_b,\hat{H}_b]+\gamma \hat{\mathcal{D}}\rho_b+\kappa\rho_a-\kappa\rho_b  \label{eq:Master_dis}
 \end{eqnarray}
where $\hat{\mathcal{D}}\rho=\sum_i n_i\rho n_i-\frac 12 (n^2_i\rho+\rho n^2_i)$ represents the effect of the white noise. Since it is still a translational invariant and noninteracting system,  the dynamics of $P(t)$ can be easily solved based on Eq.~(\ref{eq:Master_dis}). As shown in Fig.~\ref{fig:Supp}(f), we can find that even though the external white noise makes the system decay faster since it facilitates the heating, the long-time behavior is still qualitatively the same as that in the noise-free case $P(t)\sim t^{-0.5}$. This indicates that the white noise is an irrelevant perturbation for the long-time behavior.

%







\end{document}